# Experimental Quantification of Spin–Phonon Coupling in Molecular Qubits using Inelastic Neutron Scattering

Stefan H. Lohaus[1], Kay T. Xia[1], Yongqiang Cheng[2], Ryan G. Hadt[1,*]

[1]Division of Chemistry and Chemical Engineering, California Institute of Technology, Pasadena, CA 91125.
[2]Neutron Scattering Division, Oak Ridge National Laboratory, Oak Ridge, TN 37831.

*Corresponding author: rghadt@caltech.edu

## Abstract

Electronic spin superposition states enable nanoscale sensing through their sensitivity to the local environment, yet their sensitivity to vibrational motion also limits their coherence times. In molecular spin systems, chemical tunability and atomic-scale resolution are accompanied by a dense, thermally accessible phonon spectrum that introduces efficient spin relaxation pathways. Despite extensive theoretical work, there is little experimental consensus on which vibrational energies dominate spin relaxation or how molecular structure controls spin–phonon coupling (SPC). We present a fully experimental method to quantify SPC coefficients by combining temperature-dependent vibrational spectra from inelastic neutron scattering with spin relaxation rates measured by electron paramagnetic resonance. We apply this framework to two model $S = 1/2$ systems, copper(II) phthalocyanine (CuPc) and copper(II) octaethylporphyrin (CuOEP). Two distinct relaxation regimes emerge: below 40 K, weakly coupled lattice modes below 50 cm$^{-1}$ dominate, whereas above 40 K, optical phonons above ~185 cm$^{-1}$ become thermally populated and drive relaxation with SPC coefficients nearly three orders of magnitude larger. Structural distortions in CuOEP that break planar symmetry soften the crystal lattice and enhance anharmonic scattering, but also raise the energy of stretching modes at the molecular core where the spins reside. This redistributes vibrational energy toward the molecular periphery and out of plane, ultimately reducing SPC relative to CuPc and enabling room-temperature spin coherence in CuOEP. Although our method does not provide mode-specific SPC coefficients, it quantifies contributions from distinct spectral regions and establishes a broadly applicable, fully experimental link between crystal structure, lattice dynamics, and spin relaxation.

## Introduction

The application of quantum bits (qubits) as nanoscale sensors has enabled measurements at unprecedented spatial and sensitivity scales, from detecting single-neuron action potentials (1) and mapping subcellular temperature variations (2), to single-protein NMR spectroscopy (3). These experiments were achieved by creating superpositions of spin states and exploiting their high sensitivity to the environment. While record room-temperature coherence times of up to 1.8 ms were achieved for spins housed at nitrogen-vacancy defects in a diamond lattice (4), their bulky carbon framework restricts chemical tunability and limits the spatial resolution to several nanometers. An alternative molecular approach uses unpaired electron spins of paramagnetic complexes. Besides allowing for further miniaturization of quantum sensors, molecules are compatible with a range of physical and biological environments (5), and have highly tunable structures, enabling chemists to adjust their electronic transitions and vibrational spectra (6–8).

For quantum sensors, spins must respond sensitively to external signals while remaining well shielded from environmental noise to preserve long coherence times (9). This has been achieved by minimizing nearby electronic and nuclear spins that induce decoherence through dipole–dipole and hyperfine interactions (4, 10). Interactions with lattice vibrations, however, are unavoidable and become increasingly significant as more phonon modes are thermally populated. At room temperature, spin–phonon coupling (SPC) becomes the dominant relaxation channel, limiting coherence and destroying spin superposition (11). As discussed below, many recent studies have modelled spin relaxation using *ab initio* and ligand-field frameworks to pinpoint the most strongly coupled modes. Yet, predictions remain inconsistent, and experimental quantification of individual coupling coefficients has not been achieved. Even phonon data are rare for molecular qubits, as most experimental efforts focus on spin dynamics.

Here, we take a fully experimental approach, measuring vibrational spectra by inelastic neutron scattering (INS) and correlating them to the spin relaxation determined by pulse electron paramagnetic resonance (EPR). We chose copper(II) phthalocyanine (CuPc) and copper(II) octaethylporphyrin (CuOEP), two model systems with one unpaired electron at their Cu(II) centers ($S = 1/2$), see Fig. 1a,b. When placed in a magnetic field, the $S = 1/2$ Zeeman sublevels split, forming a two-level system. Coherent spin superpositions can be generated by microwave pulses in an EPR spectrometer that excites the spin to its higher-lying state. Relaxation through interactions with lattice vibrations destroys this superposition, with a characteristic timescale defined as the spin–lattice relaxation time $T_1$.

Many studies have sought to identify which phonon modes most strongly contribute to spin relaxation in $S = 1/2$ systems. Ligand-field models predict that high-energy optical modes involving symmetric metal–ligand stretching possess the symmetry required to induce the largest variations in the electron's Zeeman splitting ($g$ value) (12–14). SPC arises from spin–orbit–coupling–mediated mixing between ligand-field ground and excited states, which is modulated by lattice vibrations. Raman spectroscopy has indeed linked such symmetric stretches (200–300 cm$^{-1}$) to the temperature dependence of $T_1$ above 40 K in Cu(II) porphyrins (15). When molecular symmetry is lowered, this selection rule relaxes, enabling additional vibrational modes to couple linearly with the spin (14).

Beyond these symmetry-based models, *ab initio* calculations remain divided and at times contradictory about which vibrations govern relaxation. One study reports comparable coupling strengths across both low- and high-energy modes (16); another attributes the strongest effects to modes above 200 cm$^{-1}$ (17), while others find that only the lowest optical phonons (< 50 cm$^{-1}$) contribute significantly (18, 19). Single-crystal EPR measurements on Cu(acac)$_2$ show that both high-energy optical and low-energy lattice vibrations contribute to spin relaxation at different temperatures. The data reveal a rotation of the relaxation tensor axes as temperature decreases, marking a transition in the dominant SPC mechanism, from intramolecular optical vibrations to collective lattice modes at low temperatures (20).

Another key factor is the mixing between vibrational modes of optical and acoustic character. Acoustic modes with long wavelengths barely distort individual molecules and cannot directly couple to spins (21). However, they can become active in spin relaxation when mixed with intramolecular modes (22, 23). This understanding has led to design strategies that reduce the number of low-energy modes and use rigid ligands to reduce the motion of the molecular core (23–25), as well as combining high-energy intramolecular modes with weak intermolecular interactions to minimize coupling between optical and acoustic modes (26).

Although phonons lie at the heart of spin–lattice relaxation, direct experimental observation of their spectra in $S = 1/2$ molecules remains remarkably scarce. Optical spectroscopies (Raman, IR, THz) are bound to selection rules and access only zone-centers, while phonon dispersions have been experimentally mapped in only two systems: VO(acac)$_2$ by INS (22) and VO(TPP) by inelastic X-ray scattering (IXS) (27). They highlighted the importance of low-energy modes in spin relaxation, revealing optical vibrations as low as 10 cm$^{-1}$ due to the softness of the porphyrin lattice. Both techniques are time-consuming and involve major experimental challenges and complex data reduction, hindering systematic, temperature-dependent investigations across molecular systems. For instance, single-crystal INS requires large, well-aligned molecular crystals and laborious rotation to probe different orientations, while compounds are often susceptible to X-ray radiation damage in IXS experiments.

Using an alternative instrument design, we employ INS to directly probe the full vibrational spectrum of $S = 1/2$ complexes up to 8000 cm$^{-1}$ (the highest-energy phonon feature appears at 3100 cm$^{-1}$). The inverted geometry of the VISION spectrometer at the Oak Ridge National Laboratory (28, 29) integrates over momentum, sacrificing momentum resolution in exchange for rapid, high-throughput acquisition of complete spectra. Each spectrum was measured in under 30 minutes using ~500 mg of powder. This capability enables temperature-dependent measurements between 5–300 K, capturing anharmonicity through line broadening and frequency shifts. By combining these vibrational spectra with pulse EPR measurements of spin-lattice relaxation, we propose a fully experimental framework to interpret SPC. This approach identifies the spectral regions most relevant to spin relaxation and provides, for the first time, experimental SPC coefficients for low- and high-energy modes, bridging the gap between theory-driven models and experiment.

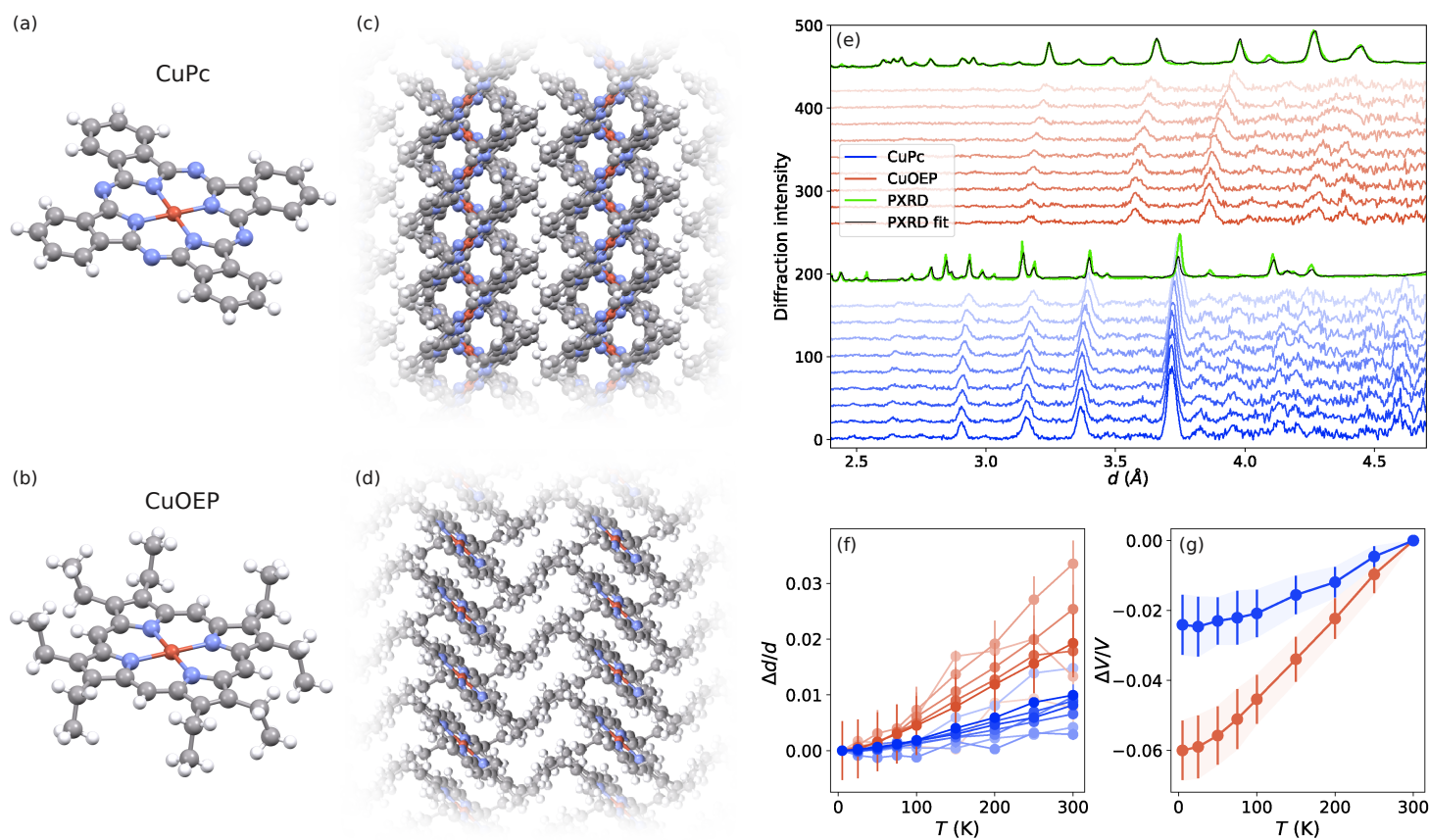

*Figure 1.* Structural characterization. Molecular (a,b) and crystal structures (c,d) of CuPc and CuOEP. (e) Neutron diffraction patterns collected between 5–300 K (dark to light shades) compared to PXRD at 300 K. (f) Temperature-dependent d-spacing changes along different crystallographic directions (different shades), obtained from fits to the peaks in panel (e), with error bars representing one-sigma fitting uncertainties. (g) Fractional volume change assuming an isotropic lattice expansion (see SI Section 2). Errors indicate the standard deviation between the isotropic prediction and the measured anisotropic d-spacings.

## Molecular and crystal structures

CuPc is a planar molecule with four coordinating nitrogens and a ligand of fused pyrrole and benzene rings, forming a $D_{4h}$ framework (Fig. 1a). CuOEP has the same $CuN_4$ coordination, but the peripheral benzene rings are opened into β-ethyl groups that protrude out of the plane (Fig. 1b). This results in a saddled macrocycle with a square planar core (15, 30). The Cu–N bonds are slightly shorter for CuPc (1.934 Å (31)) than CuOEP (1.998 Å (32)).

These structural differences result in different crystals (Figs. 1c,d): CuPc is monoclinic while CuOEP is triclinic. The most stable polymorph of CuPc (β-phase) crystallizes with two molecules in the unit cell, nearly orthogonal to each other (91.6°). Molecules form a herringbone (criss-cross) stack along the b direction, 3.34 Å apart, with a lateral offset where the Cu stacks over the outer aza (bridge) N atoms of the molecule below (31). The saddled β-ethyl groups of CuOEP inhibit the formation of a herringbone structure. Instead, there is a single molecule per unit cell, and all molecules are stacked, with Cu atoms over the pyrrole rings (30, 32). Instead of having an orthogonal molecule in the crystal, like CuPc, the interaction across layers is facilitated by the out-of-plane β-ethyl groups and by π-π interactions. The mean orthogonal distance between molecular layers is 3.33 Å, virtually the same as in CuPc.

The lattice structures of CuPc and CuOEP were characterized by X-ray and neutron diffraction (Fig. 1e). Their distinct powder X-ray diffraction (PXRD) patterns (green curves) were indexed and Rietveld-refined against the corresponding crystal structures shown in Fig. 1c,d using GSAS-II (black curves). *In situ* neutron diffraction enabled quantification of thermal expansion for both compounds, although only a few Bragg reflections are visible due to the strong incoherent scattering cross-section of hydrogen. The positions of individual peaks in *d*-space were tracked by Gaussian fits (Fig. 1f). The change in lattice volume was determined, assuming isotropic thermal expansion, using PXRD values measured at room temperature as references (Fig. 1g). For CuPc, a full monoclinic fit yielded a volume expansion equivalent to the isotropic model (see SI Section 2). CuOEP exhibits more than twice the expansion of CuPc, reflecting its softer lattice arising from the out-of-plane β-ethyl groups that weaken intermolecular interactions. All fits, models, and peak indexing details are provided in Section 2 of the SI.

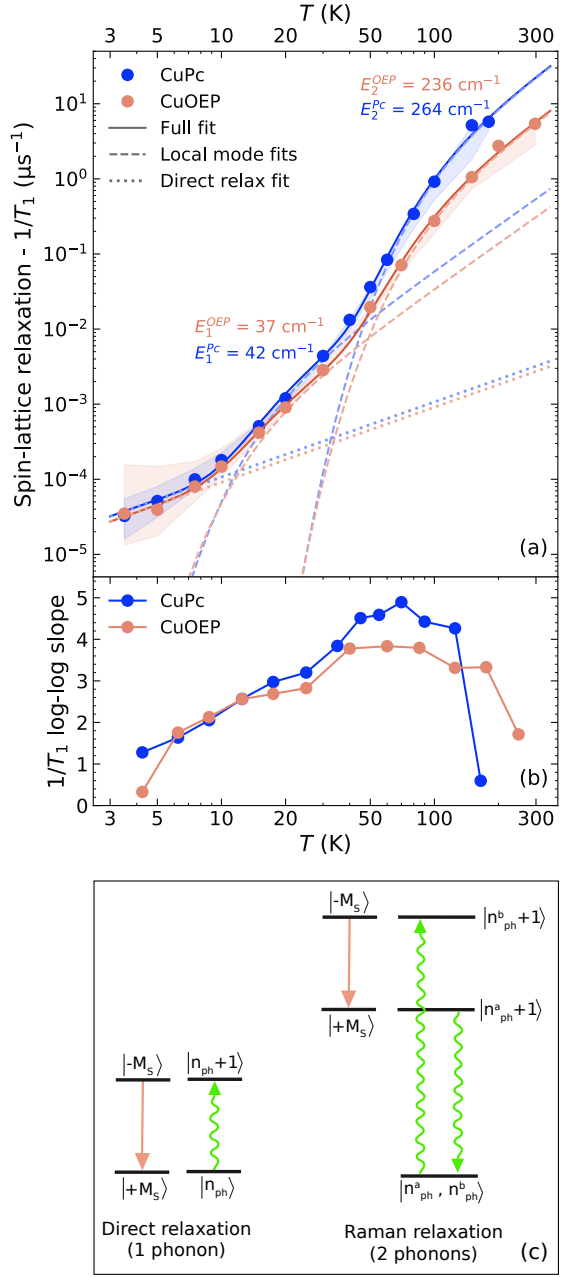

*Figure 2.* *Spin dynamics characterization. (a) Spin-lattice relaxation rates ($1/T_1$) of CuPc and CuOEP measured by pulse EPR at perpendicular field orientations. Data were collected by saturation recovery below 30 K and inversion recovery above. The shaded areas indicate the full range of data acquired at parallel and perpendicular field orientations, with both inversion and saturation recovery experiments (see Methods and SI Section 3). Dashed lines show local-mode fits (Eq. 1), dotted lines the one-phonon term, and solid curves the combined fit. Dashed lines represent local-mode fits using Eq. 1, dotted lines show the one-phonon contributions, and solid curves the combined fit. (b) Logarithmic slopes of the data from panel (a), giving the power-law exponent in the temperature dependence of $1/T_1$. (c) Schematic illustration of spin relaxation mechanisms involving one-phonon (direct) and two-phonon (Raman) processes.*

## Spin-lattice relaxation

Spin–lattice relaxation rates ($1/T_1$) of CuPc and CuOEP, measured by pulse EPR between 3.5–300 K, are shown in Fig. 2a. To minimize spin–spin interactions and isolate intrinsic spin–lattice relaxation, powder samples were diluted into isostructural diamagnetic matrices (ZnPc and ZnOEP) at a 1:1000 molar ratio. Metal substitution from Cu to Zn leads to only small modifications of the lattice parameters, with maximum relative changes of 1.24% for Pc and 2.41% for OEP (see SI Section 1). EPR data were collected using inversion- and saturation-recovery sequences at different magnetic field positions corresponding to molecular orientations parallel and perpendicular to the applied field. Spectral-diffusion effects, which are minimized in saturation-recovery measurements, are noticeable only at low temperatures. Accordingly, Fig. 2 presents saturation-recovery data below 30 K and inversion-recovery data at higher temperatures (at the perpendicular field orientation). The complete data set is shown in Fig. 5 and discussed in Section 3 of the SI. Our measurements are consistent with previous pulse EPR studies at high temperatures (11, 15) and extend those results to lower temperatures with greater temperature resolution.

As temperature increases, phonon populations rise, accelerating spin relaxation. Both molecules exhibit comparable relaxation rates at low temperature, but at elevated temperatures CuPc relaxes more rapidly. By 180 K, its excited-spin lifetime becomes too short to be detected by pulse EPR, whereas CuOEP remains coherent up to room temperature ($T_1$ for CuOEP at 300 K was taken from the 1:100 dilution data (15), as the 1:1000 sample signal was too weak). Changes in the slope of $1/T_1(T)$ indicate the onset of additional relaxation mechanisms (Fig. 2b shows the logarithmic derivative of panel a). Two such transitions appear near 7 K and 40 K. Below 7 K, the nearly linear dependence ($1/T_1 \propto T$) corresponds to the *direct* process, where a spin relaxes by emitting a single phonon of matching energy (see Fig. 2c). Both CuPc and CuOEP exhibit comparable contributions from this process. The direct pathway is inefficient, however, because the phonon density is extremely low at the spin-transition energy (~0.14 cm$^{-1}$ at 0.3 T).

Two-phonon Raman processes dominate spin relaxation above 7 K. In these processes, one phonon is absorbed while another is emitted to mediate the spin transition (Fig. 2c). Only the energy difference between the two phonons must match the spin splitting, allowing the entire thermally accessible phonon spectrum to contribute. Because one phonon must be thermally excited, the onset temperature of Raman relaxation reflects the energies of the participating phonon modes. To extract these characteristic energies, we fit the data using a *local-mode model*, which explicitly considers the probabilities of two-phonon processes. Phonons are bosonic quasiparticles whose occupation numbers follow the Bose–Einstein distribution, $n(E,T) = 1/(e^{E/k_B T} - 1)$. For Raman relaxation, one phonon is thermally populated and absorbed while another is created. Assuming these two phonons have the same energy (since the spin-transition energy is much smaller than the phonon energies), the relaxation probability becomes (19):

$$\left(\frac{1}{T_1}\right)_{2-ph} \propto n(E_{ph}, T) \cdot (n(E_{ph}, T) + 1) = \frac{e^{E_{ph}/k_B T}}{(e^{E_{ph}/k_B T} - 1)^2} \qquad (1)$$

This function has the same $T^2$ high-temperature limit as the widely used Debye model. In contrast, the local-mode model allows direct extraction of the phonon energies involved in spin relaxation and remains applicable across all energies beyond the Debye (quadratic) phonon regime, which extends only up to ~15 cm$^{-1}$ in molecular systems. Alternative Debye-model fits are provided in Section 3 of the SI.

The relaxation data in Fig. 2a were fitted using two local modes in addition to the linear direct process. Relaxation between 7–40 K is mediated by phonons around 42.5 ± 5.7 cm$^{-1}$ for CuPc and 37.1 ± 4.8 cm$^{-1}$ for CuOEP, while high-temperature relaxation involves modes around 264.8 ± 14.4 cm$^{-1}$ and 236.2 ± 12.9 cm$^{-1}$, respectively (errors denote one-sigma fit uncertainties). Without direct phonon measurements, this is the full extent of vibrational information that can be inferred from spin-relaxation fits. In the following section, we show how experimental phonon data provide crucial insights into spin–lattice relaxation and enable quantification of coupling strengths.

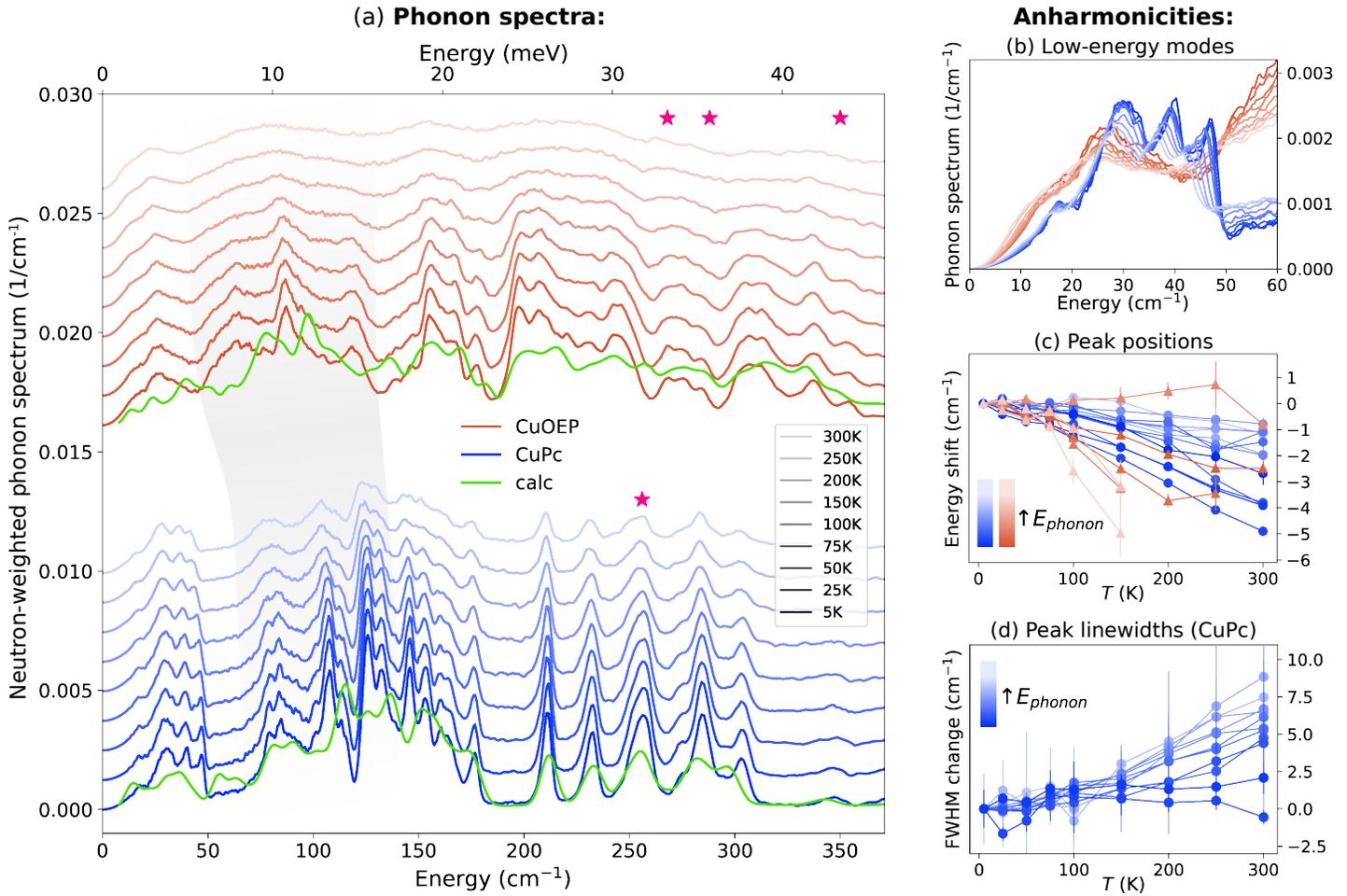

*Figure 3.* Phonon behavior of CuPc and CuOEP. (a) Normalized phonon spectra measured by INS between 5–300 K (after corrections described in Methods and SI Section 4). Green curves show DFT-calculated phonon spectra. Stars mark the calculated energy of the symmetric stretch of CuPc and the three modes of CuOEP with the largest symmetric stretching character. Gray shading visualizes the redshifted spectrum of CuOEP compared to CuPc. (b) Expanded view of the low-energy phonon region. (c) Anharmonic energy shifts of CuPc (blue) and CuOEP (orange), with lighter shades corresponding to higher phonon energies. Error bars show one-sigma peak fitting uncertainties. (d) Temperature-dependent phonon broadening of CuPc, quantified as changes in FWHM from Gaussian fits.

## Phonon spectra

We employ INS to measure the full vibrational spectra of CuPc and CuOEP between 5–300 K, with most of the thermally accessible region shown in Fig. 3a. VISION measures the self-dynamic structure factor $S(Q,E)$ along a fixed $Q$–$E$ trajectory, which prevents a strict extraction of the phonon DOS without computational input (details in SI Section 4.6). The measured neutron scattering intensities reflect the same vibrational excitations, but with slightly different spectral weights than the DOS. Because neutrons interact with different atomic species according to their scattering cross sections, the spectra of organic complexes are often dominated by hydrogen motion. In this thermally accessible region, however, the vibrations involve collective motions of all atoms, such that the data presented here represent the full vibrational spectrum, with intensities weighted by hydrogen participation.

To minimize computational input, we present here the neutron-weighted VISION spectra and discuss the effects of using the phonon DOS in the SPC analysis below. The spectra were corrected for excitation probabilities (Bose correction), multiphonon excitations, and instrumental background (see SI Section 4 for details and the full dataset). All spectra are normalized to unity up to 600 cm$^{-1}$, which is taken here as the threshold of thermal accessibility at 300 K. The DFT-calculated phonon spectra (see Methods), shown in green, capture most of the measured features and are used to assign the character of the observed modes (*vide infra*).

CuPc has sharp phonon modes up to room temperature and a distinct set of low-energy translational and bending modes up to about 50 cm$^{-1}$, with the lowest peak detected at 17 cm$^{-1}$. These modes, detailed in Fig. 3b, include acoustic lattice phonons and optical vibrations. The latter are often called *pseudo-acoustic* modes because they have a non-flat dispersion and transport energy through the crystal. Such modes between 25–60 cm$^{-1}$ have been considered main contributors for spin relaxation of $S > 1/2$ single-molecule magnets (8, 24). The quadratic energy dependence below 10–15 cm$^{-1}$ comes from the Debye model used to remove the experimental elastic line.

The next set of modes of CuPc between 70–200 cm$^{-1}$ includes higher-order bending modes (doming, saddling, ruffling) and an asymmetric stretch around 175 cm$^{-1}$ (calculated value). A sparser set of modes starts above 200 cm$^{-1}$, which includes additional higher-order bending but also in-plane modes like torsion (212 cm$^{-1}$), scissoring (229 cm$^{-1}$), and the symmetric stretch (256 cm$^{-1}$, marked with a star), which is regarded as key for spin relaxation. Since there are two molecules per unit cell, these calculated modes appear in pairs with nearly degenerate frequencies due to slight structural asymmetries. The mode assignments for both CuPc and CuOEP agree with previous calculations and resonance Raman measurements (14, 15, 33, 34).

CuOEP has a similar phonon spectrum, but with significantly broader linewidths and a nearly featureless spectrum at room temperature. Most of its spectrum is shifted to lower energies compared to CuPc. While modes below 50 cm$^{-1}$ are redshifted by about 5 cm$^{-1}$ with respect to CuPc, modes between 50–130 cm$^{-1}$ have more significant shifts of up to 20 cm$^{-1}$ (visualized by the gray band in Fig. 3a). These correspond to different bending modes, which are particularly affected by the peripheral softness of the β-ethyl groups compared to the planar benzene rings of CuPc.

Another striking difference emerges from the calculations: the normal modes of CuOEP exhibit strongly mixed vibrational character, combining in- and out-of-plane motion (animations are provided as an SI). In CuPc, the $D_{4h}$ symmetry keeps stretching and out-of-plane modes largely separable, but in CuOEP the saddled macrocycle (~$D_{2d}$) lifts these degeneracies, allowing eigenvectors to become linear combinations of motions that were previously symmetry-distinct. As a result, CuOEP shows multiple mixed stretches with substantial out-of-plane motion, rather than a single characteristic stretching frequency, consistent with prior DFT and resonance Raman studies (15). In contrast to the lower-energy modes, which are redshifted relative to CuPc, the dominant CuOEP stretching modes occur at higher energies (268, 288, and 350 cm⁻¹; marked with stars in Fig. 3a). This indicates stiffer bonds at the CuOEP molecular core compared to CuPc, alongside softer out-of-plane modes arising from its molecular geometry.

## Phonon anharmonicities

At low temperatures, atomic displacements are small and vibrations are well described by harmonic potentials, where phonons behave as independent oscillators with infinite lifetimes and fixed frequencies. The *quasiharmonic* model adds the effect of thermal expansion: as the lattice volume increases, restoring forces weaken and phonon energies soften. The potential remains quadratic, but mode frequencies are renormalized through their Grüneisen parameters $\gamma_i$:

$$\frac{\Delta E_i}{E_i} = \gamma_i \frac{\Delta V}{V} \qquad (2)$$

Beyond this approximation, cubic and quartic terms in the potential enable energy exchange between modes and couple vibrations to electronic or spin degrees of freedom. These *pure* anharmonic interactions shorten phonon lifetimes and broaden spectral features, while the cubic term's asymmetry further softens frequencies at large amplitudes.

For CuPc and CuOEP, the temperature-dependent softening of phonon energies and broadening of linewidths indicate non-harmonic behavior. To quantify these effects, phonon peaks were fitted with Gaussian and Voigt profiles (see SI Section 6 for details). In CuPc (Fig. 3c,d), low-energy modes (<150 cm⁻¹) soften by up to ~5 cm⁻¹ at 300 K with negligible linewidth changes, whereas higher-energy optical modes (>200 cm⁻¹) remain nearly fixed in energy but broaden significantly, signaling shorter lifetimes. In CuOEP, strong spectral overlap hinders reliable fits to individual modes, so trends are primarily qualitative. However, both low- and high-energy modes show pronounced softening (particularly the feature near 240 cm⁻¹), suggesting stronger anharmonicity.

Thermal expansion from *in situ* neutron diffraction (Fig. 1g) was used to model these frequency shifts via the quasiharmonic relation (Eq. 2). In both systems, the softening of low-energy modes scales approximately linearly with lattice expansion, indicating that Grüneisen parameters $\gamma_i$ capture most temperature-dependent behavior (see Figs. S42, S45). For low-energy modes, $\gamma_i$ values are significantly larger for CuPc ($\gamma_i \approx 4$) than CuOEP ($\gamma_i \approx 2$). Since CuPc and CuOEP show comparable absolute frequency shifts at these energies, but CuOEP expands considerably more, its vibrational spectrum is less sensitive to lattice expansion. The β-ethyl substituents of CuOEP weaken intermolecular interactions, softening the lattice but also partially decouple its vibrational modes from lattice strain. Such vibrational isolation of molecules in the CuOEP crystal could contribute to its reduced $1/T_1$ compared to CuPc.

In CuPc, the trend of softening without broadening at low energies suggests that these phonons are well described by a quasiharmonic lattice expansion model, while the line broadening of its high-energy optical modes implies reduced lifetimes from pure anharmonic scattering. The spectra of CuOEP are considerably broader than CuPc. Both samples formed good crystals (Fig. 1e), so we discard disorder-induced broadening. The lowering of symmetry can reduce phonon lifetimes by opening additional scattering channels otherwise forbidden in higher-symmetry structures (35, 36). The structural distortions and mixed character of CuOEP's eigenvectors could activate such additional phonon-phonon processes.

This would also explain the enhanced energy softening of its high-energy modes compared to CuPc. Mixing of localized stretching modes with out-of-plane motion could transfer the typically stronger volume dependence of bending modes (higher $\gamma_i$) to stretching ones.

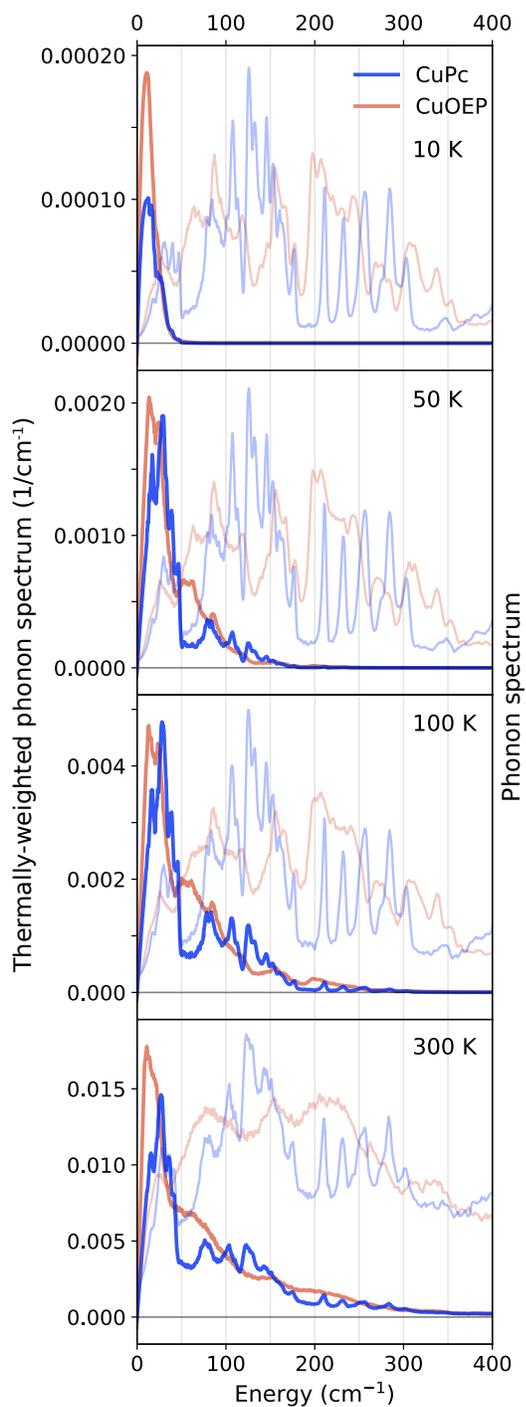

**Figure 4**. Thermal phonon population of CuPc and CuOEP. Bose-weighted phonon spectra at different temperatures, overlaid with their measured phonon spectra for reference (right axis, light-shaded curves). The spectra at 10 K are linear interpolations between 5–25 K measurements.

## Thermal phonon population

Beyond their vibrational energies and anharmonicities, understanding phonon behavior requires considering their thermal population. This is relevant for spin–lattice relaxation, as the Raman process depends on the prior occupation of phonon modes. The phonon mode occupation at each temperature can be visualized by weighting the phonon spectrum by the Bose–Einstein occupation $n(E,T)$, as shown in Fig. 4. Even though CuOEP has a lower phonon density below 50 cm⁻¹, its lower vibrational energies compared to CuPc result in a larger population below 10 K. At this temperature, modes around 40 cm⁻¹ become thermally accessible, consistent with the local-mode fits in Fig. 2a, which assigned SPC to modes near 42 cm⁻¹ and 37 cm⁻¹ for CuPc and CuOEP, respectively. By 40 K, all modes below 200 cm⁻¹ are populated. However, the absence of new relaxation channels between 10–40 K suggests that modes in this range contribute little to spin relaxation. Beyond 40 K, phonons above 200 cm⁻¹ become accessible. Although their populations remain smaller than those of low-energy modes, they dominate relaxation at higher temperatures, implying stronger coupling strengths.

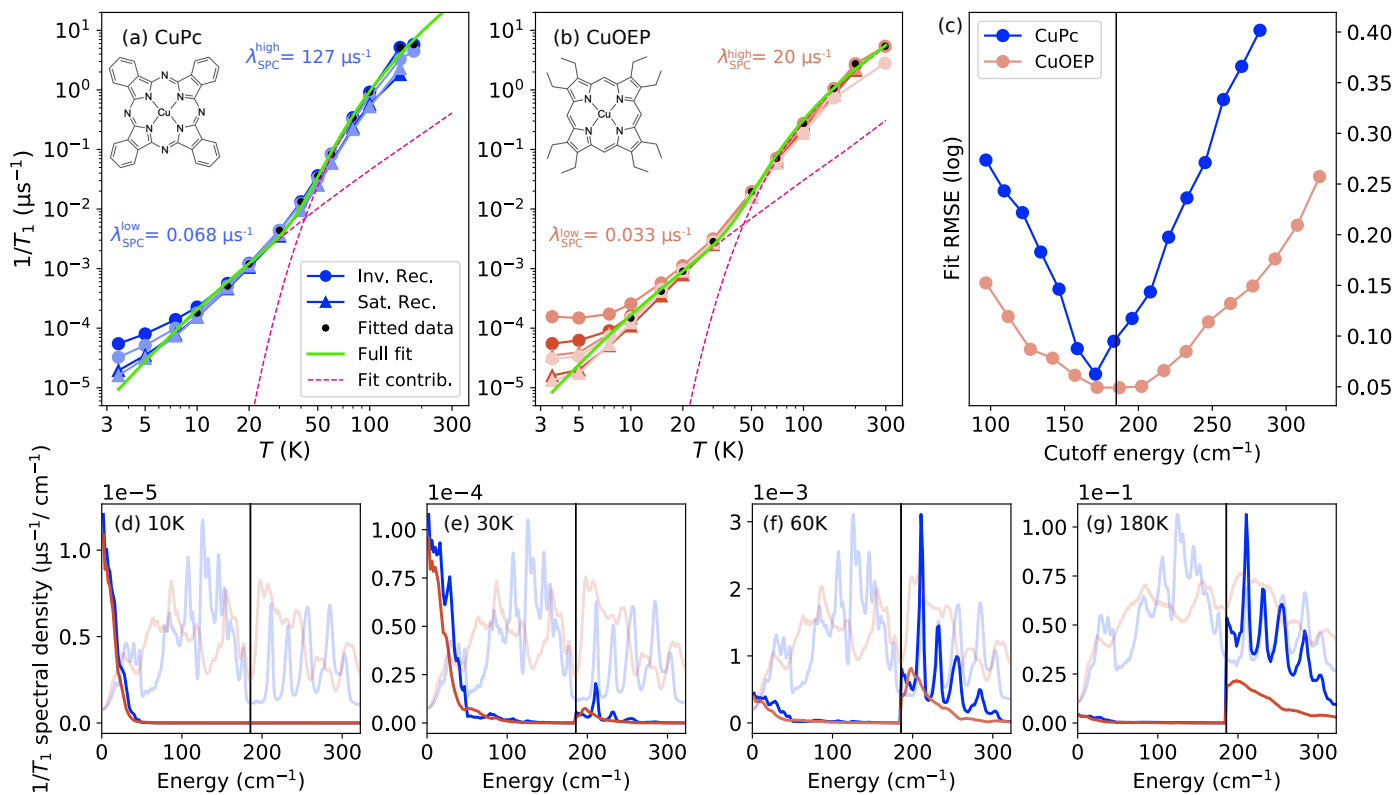

*Figure 5.* Spin–phonon coupling in CuPc and CuOEP. (a,b) Spin–lattice relaxation of CuPc and CuOEP measured by pulse EPR using inversion- and saturation-recovery sequences. Different shades represent measurements at different magnetic fields. Green curves show fits to Eq. 3, while pink dashed curves indicate the contributions from low- and high-energy phonons. (c) Root-mean-square fitting errors, computed on a logarithmic scale, as a function of the cutoff energy separating low- and high-energy phonon regions. (d–g) Spin–lattice relaxation spectral densities at various temperatures, visualizing the parts of the phonon spectrum that participate in the relaxation process.

## Spin-phonon coupling

To go beyond a qualitative discussion of SPC, we expand the local mode fitting model of Eq. 1. By including the measured phonon spectra, $G_T(E)$, we can integrate over all energies, and instead of fitting phonon energies, we extract their coupling strengths $\lambda_{SPC}(E)$. The two-phonon Raman relaxation rate $1/T_1$ at a given temperature $T$ is given by:

$$\left(\frac{1}{T_1}\right)_T = \int_0^{E_{ph,max}} \lambda_{SPC}(\varepsilon) \cdot G_T(\varepsilon) \cdot \frac{e^{\varepsilon/k_B T}}{(e^{\varepsilon/k_B T} - 1)^2} d\varepsilon, \qquad (3)$$

where $G_T(E)$ is normalized to one and $E_{ph,max} = 600\ cm^{-1}$. We assume $\lambda_{SPC}(E)$ to be temperature-independent (it is, at most, weakly dependent on temperature (17)). To avoid overfitting, we do not include direct relaxation and only use data above 10 K for the fits.

The main goal is to compare the contributions of highly populated low-energy vibrations to those of strongly coupled optical modes. Thus, we divide the coupling coefficient into low- and high-energy regions: $\lambda_{SPC}(E) = \lambda_{SPC}^{low} + \lambda_{SPC}^{high}$. Figs. 5a-b show the fits of Eq. 3 with these two coefficients, giving $\lambda_{SPC}^{low}$ = 0.068 µs$^{-1}$ and $\lambda_{SPC}^{high}$ = 127 µs$^{-1}$ for CuPc, and $\lambda_{SPC}^{low}$ = 0.033 µs$^{-1}$ and $\lambda_{SPC}^{high}$ = 20 µs$^{-1}$ for CuOEP. Both molecules have low-energy couplings that are three orders of magnitude lower than the high-energy coefficients. The effect of field position (perpendicular vs. parallel) and method (inversion-recovery vs. saturation-recovery) on the fitted parameters is discussed in Section 7 of the SI.

The cutoff energy separating the low- and high-energy regions was determined by minimizing the logarithmic root-mean-squared error of the fits (Fig. 5c). A value of 185 cm$^{-1}$ lies near a common minimum and coincides with a low-intensity spectral gap for both CuPc and CuOEP, providing a natural boundary between the two regimes. The temperature crossover between these energy regions occurs at 42 K for CuPc and 45 K for CuOEP. The latter closely matches the 47 K crossover identified by pulse EPR anisotropy measurements on CuOEP (20), which marks the transition from lattice-dominated to more localized relaxation processes.

To test the robustness of our method, we evaluated the SPC coefficients using different normalization cutoff energies and phonon representations. Table 1 summarizes the main results, with full details in Section 7 of the SI. Lowering the normalization cutoff energy from 600 to 380 cm$^{-1}$ enhances the spectral intensities and correspondingly reduces the extracted SPC coefficients by 15% for CuOEP and about 30% for CuPc. The ratio between $\lambda_{SPC}^{high}$ and $\lambda_{SPC}^{low}$ remains mostly unchanged, indicating that the relative strengths are preserved. Using the phonon DOS (with computational input) instead of the VISION spectra affects the extracted parameters more strongly. In particular, $\lambda_{SPC}^{low}$ roughly doubles, whereas $\lambda_{SPC}^{high}$ changes by less than 30%, reflecting the redistribution of spectral weight toward higher energies in the DOS. In all cases, $\lambda_{SPC}^{high}$ is nearly three orders of magnitude larger than $\lambda_{SPC}^{low}$.

*Table 1. Robustness of the extracted SPC coefficients (in µs⁻¹) against variations in normalization cutoff energy and phonon representation. The reference corresponds to fits of Fig. 5.*

| Analysis condition | Phonon spectrum | $E_{cutoff}$ (cm⁻¹) | CuPc | | CuOEP | |
|---|---|---|---|---|---|---|
| | | | $\lambda_{SPC}^{low}$ | $\lambda_{SPC}^{high}$ | $\lambda_{SPC}^{high}$ | $\lambda_{SPC}^{high}$ |
| Reference | VISION | 600 | 0.068 | 127 | 0.033 | 20 |
| Normalization | VISION | 380 | 0.045 | 95 | 0.028 | 17 |
| Phonon representation | DOS | 600 | 0.135 | 111 | 0.082 | 26 |

This method allows us to visualize the spectral regions that participate in spin relaxation. Figs. 5d-g show the spectral contributions to $1/T_1$ (integrand of Eq. 3) at different temperatures, overlaid with their phonon spectra for reference (light-shaded curves). Low-energy modes below 50 cm⁻¹ dominate the relaxation at low temperatures. The contribution from the strongly-coupled optical modes above 185 cm⁻¹ becomes evident already at 30 K and grows quickly at higher temperatures. Interestingly, vibrational modes between 50–185 cm⁻¹ contribute weakly to spin relaxation across all temperatures. This is due to their lower population compared to lower-energy modes and their weaker coupling strength compared to higher-energy phonons. This explains why the redshifted spectrum of CuOEP compared to CuPc in this region (gray shaded area in Fig. 3a) has little effect on its spin relaxation. An alternative fit, that introduces a third energy window to isolate this spectral region, gives the same behavior: fitted coefficients change slightly, but modes between 50–185 cm⁻¹ still contribute significantly less to spin relaxation than lower- and higher-energy vibrational modes (see Figs. S52 and S53).

In the low-energy/low-temperature regime, spin–lattice relaxation follows a thermodynamic behavior: all phonon modes are weakly coupled to the spin, and their contribution depends solely on their thermal population. Although CuOEP exhibits a more populated phonon spectrum below 50 cm⁻¹ (Fig. 4a, 10 K), its coupling strength is roughly half that of CuPc (0.068 vs. 0.033 µs⁻¹), resulting in slightly slower relaxation. Pure acoustic modes, dominated by rigid molecular translations, are not expected to couple with spins (21), and if such modes below 15 cm⁻¹ are excluded from the fit, the difference in coupling strength between CuPc and CuOEP persists (the absolute SPC values increase to 0.135 and 0.084 cm⁻¹ due to the large thermal factor of such modes, see Fig. S51).

As temperature increases, spin relaxation deviates from the thermally-driven regime, and modes above 185 cm⁻¹ dominate. In both molecules, these strongly-coupled regions comprise in-plane scissoring modes, torsions, and symmetric stretches, as well as higher-order out-of-plane ruffling and saddling. Here, the coupling in CuPc is 6.4 times stronger than in CuOEP (127 vs. 20 µs⁻¹). Ligand-field models for spin relaxation have predicted that symmetric stretches produce the largest modulations of the ground-state $g$ value and minority spin mixings and, thus, contribute strongly to $T_1$ (12–14). However, the fitted cutoff energy at 185 cm⁻¹ indicates that additional modes with energies below the symmetric stretch are required to reproduce the data.

In ligand-field theory, the energy separation to electronic excited states determines how strongly vibrations can modulate spin–orbit coupling and hence the SPC strength (13, 14). However, the stronger coupling in CuPc compared to CuOEP cannot be attributed to their ligand-field transition energies, which are marginally smaller for CuOEP (15, 37). Calculations have also shown that broader phonon linewidths can correlate with weaker Raman relaxation in single-molecule magnets (38), although CuOEP exhibits slower relaxation despite its broader modes, suggesting that linewidth effects are not the dominant factor here.

We attribute CuOEP's weaker SPC to a combination of its weaker intermolecular interactions and the increased stiffness of its first coordination sphere. This stiffening, reflected in higher-energy symmetric Cu–N stretching modes, effectively isolates the localized (optical) modes that couple most strongly to spins from collective lattice motions, consistent with previously proposed vibrational decoupling strategies (23–26). We note that in our case, the suppression of low-energy phonons is less significant than this decoupling effect, since CuOEP exhibits a higher phonon density below 20 cm⁻¹. The smaller computed Grüneisen parameters for CuOEP further support this interpretation, indicating that its low-energy phonons are less affected by collective lattice expansion than those of CuPc. The out-of-plane β-ethyl substituents seem to act as vibrational dampers, reducing coupling between the molecular core and the surrounding lattice.

Phonon calculations confirm this trend through root-mean-squared displacements (RMSDs), which quantify the thermally-averaged vibrational amplitudes of individual atoms (see SI Section 8 for details). The per-atom RMSDs, obtained by summing contributions from all vibrational modes up to 400 cm⁻¹ at 300 K, are larger for CuOEP (0.32 Å) than for CuPc (0.24 Å). In contrast, the per-atom RMSDs of the molecular core (Cu and first-coordination N atoms) are smaller for CuOEP (0.15 Å) than for CuPc (0.17 Å).

Although molecular symmetry breaking can increase the number of phonon modes that couple to spins (14), the mixing of in-plane and out-of-plane stretching modes in CuOEP appears to reduce their SPC strength. Using the calculated eigenvectors, we quantified the amount of symmetric-stretching character of individual modes by considering the in-plane relative displacement amplitudes of the Cu–N bonds together with their relative phase. While all significant symmetric stretches of CuPc are concentrated near 256 cm⁻¹, CuOEP exhibits many modes with comparable but weaker stretching character, with the highest-ranked modes at 350, 268, and 288 cm⁻¹ (full ranking is provided in Section 8 of the SI). Except for the highest-energy stretch at 350 cm⁻¹, CuOEP's modes display increased out-of-plane motion and reduced in-plane stretching at the Cu–N core, resulting in a lower overall stretching character compared to CuPc. In line with this picture, reported SPC coefficients $(\partial g/\partial Q_i)^2$ indicate that CuOEP's three dominant mixed stretching modes have coupling strengths reduced by factors of 10, 1.7, and 1.4 relative to CuPc's symmetric stretch (14, 15). Transferring vibrational energy into out-of-plane motion seems to reduce the SPC efficiency. Together with the blue shift of the stretching energies, this leads to the slower spin–lattice relaxation rates seen in CuOEP.

## Conclusion

The microscopic mechanisms of spin–lattice relaxation in paramagnetic molecules remain debated, largely because direct experimental correlations between phonon spectra and spin dynamics are scarce. By combining INS with pulse EPR, we introduce the first fully experimental framework to quantify SPC coefficients in $S$ = ½ systems.

For CuPc and CuOEP, spin relaxation is mediated by two distinct regions of the vibrational spectrum. At low temperatures, relaxation occurs via weakly coupled lattice modes below 50 cm⁻¹, whereas at higher temperatures optical phonons above ~185 cm⁻¹ become thermally populated and dominate Raman relaxation with SPC coefficients nearly three orders of magnitude larger. In CuOEP, distortions that break planar symmetry soften the lattice and intermolecular interactions and redistribute vibrational motion away from the Cu–N core into peripheral and out-of-plane modes. As a result, CuOEP exhibits weaker effective spin relaxation at the Cu core across the measured temperature range compared to CuPc, despite possessing a more thermally accessible vibrational spectrum.

In CuOEP, spin–phonon decoupling emerges from a planar, rigid molecular core that preserves high-energy symmetric stretches while allowing limited out-of-plane distortions that reduce intermolecular coupling. Such a strategy must be approached with care since excessive ruffling (CuOEP → CuTPP → CuTiPP) can redshift core stretches and accelerate relaxation (15), while bulky peripheral substituents increase highly populated low-energy phonon density that can also enhance SPC (39).

Beyond CuPc and CuOEP, the presented approach is broadly applicable. It offers a relatively rapid, quantitative method to connect crystal structure, lattice dynamics, and spin relaxation without complex modeling or data analysis. The framework can extend to systems with multiple unpaired electrons, for example, where low-energy vibrations preclude spin coherence above liquid-nitrogen temperatures. Neutron spectroscopy provides access to the full vibrational spectrum, and deuteration could refine its weighting toward the thermodynamic phonon DOS. While mode-specific SPC coefficients remain inaccessible, quantifying coupling across energy ranges already yields valuable mechanistic insight. This combined INS–EPR approach provides an experimental pathway to probe structure–activity relationships and to guide the design of molecular qubits with improved coherence times, particularly near room temperature where quantum sensing applications are most relevant.

## Materials and Methods

### Synthesis

CuPc was purchased from Sigma Aldrich. CuOEP was synthesized from octaethylporphyrin ($H_2OEP$, Strem Chemicals) using a modified literature procedure (40). Both materials were characterized by PXRD and EPR before the beamtime. Synthetic details are described in Section 1 of the SI.

### Inelastic Neutron Scattering

INS is an established technique to quantify the full vibrational spectrum of materials (41). Neutrons interact directly with nuclei, without selection rules, and transfer sufficient momentum to probe the entire Brillouin zone. Measurements were performed on the VISION spectrometer (28, 29) at the Spallation Neutron Source (Oak Ridge National Laboratory), which integrates intensity over a broad range of scattering angles to yield the phonon density of states from polycrystalline samples. Approximately 500 mg of CuPc or CuOEP powder was sealed in thin-walled vanadium cans (6 mm diameter) under helium for thermal equilibration. Data were collected between 5–300 K with a white incident neutron beam, for 20 min (CuPc) and 30 min (CuOEP) per temperature, corresponding to proton charges at the target of 0.8 and 1.2 C, respectively. Data were reduced using the MANTID software package (42). Spectra were then corrected for instrumental background, Bose population, and multiphonon contributions, and the elastic line was removed before normalization in Python. Additional data analysis details are provided in Section 4 of the SI.

### Diffraction

Elastic neutron scattering data were collected simultaneously with the inelastic measurements on VISION, using time-of-flight analysis of the elastically scattered neutrons. The resulting diffraction patterns are broad for our materials due to the large incoherent cross-section of hydrogen but allowed tracking of the lattice expansion as a function of temperature while recording the phonon spectra. Complementary powder X-ray diffraction (PXRD) was performed on both materials. Measurements were carried out on a Rigaku SmartLab diffractometer with Cu Kα radiation and a Kβ filter. The diffraction patterns were Rietveld refined in GSAS-II, using the reported single-crystal structures as starting models (32, 43). Peak positions were obtained by fitting Gaussian functions. See SI Section 2 for fitting details and extracted parameters.

### Pulse electron paramagnetic resonance

Pulse EPR measures how spin polarization returns to equilibrium after excitation by short microwave pulses in a static magnetic field. Spin–lattice relaxation times ($T_1$) were determined by inversion-recovery pulse sequences at X-band using a Bruker ELEXSYS E580 spectrometer between 3.5–300 K. At low temperatures, where $T_1$ is long, spectral diffusion can contribute to the apparent relaxation times, so $T_1$ was also measured using initial saturation-recovery pulses, which suppress diffusion effects. To minimize dipole–dipole interactions and measure the effect of phonons, CuPc and CuOEP were diluted 1:1000 into isostructural diamagnetic hosts (ZnPc and ZnOEP). $T_1$ values were obtained at two field positions for CuPc (0.317 and 0.339 T) and three for CuOEP (0.313, 0.338, and 0.349 T), corresponding to parallel and perpendicular *g*-tensor orientations. Relaxation traces were fit with stretched-exponential functions to extract $T_1$ constants. Full datasets and fits are available in Section 3 of the SI.

### Phonon calculations

Spin-polarized density functional theory (DFT) calculations were performed using the Vienna *Ab initio* Simulation Package (VASP) (44) with the projector augmented-wave method (45, 46). Experimental crystal structures were used as initial models (32, 43). The CuOEP and CuPc unit cells contained 85 (one Cu) and 114 atoms (two Cu), respectively. A Hubbard U correction of 4.0 eV was applied to account for the localized Cu 3d electrons. Although magnetic coupling between Cu(II) centers is weak, antiferromagnetic spin configurations were found to slightly lower the total energy and improve numerical stability. For CuOEP, this required a 1 × 1 × 2 magnetic supercell. Electronic structure calculations employed Γ-centered k-point meshes of 2 × 2 × 3 for CuOEP and 2 × 7 × 3 for CuPc. Structures were fully relaxed, and dispersion interactions were treated using the optB86b-vdW functional (47, 48). Phonon calculations were performed using finite displacements with Phonopy (49), employing the magnetic unit cell for CuOEP and a 1 × 2 × 1 supercell for CuPc. The resulting vibrational modes were converted into simulated inelastic neutron scattering spectra using OCLIMAX (50). Additional details are available in Section 8 of the SI.

## Acknowledgements

This research used resources at the Spallation Neutron Source, a DOE Office of Science User Facility operated by the Oak Ridge National Laboratory. The beam time was allocated to VISION (BL-16B) on proposal number IPTS-32795.1. This research used computing resources made available through the VirtuES project, funded by Laboratory Directed Research and Development program and Compute and Data Environment for Science (CADES) at ORNL, as well as resources of National Energy Research Scientific Computing Center (NERSC), a U.S. Department of Energy Office of Science User Facility located at Lawrence Berkeley National Laboratory, operated under Contract No. DE-AC02-05CH11231 using NERSC award ERCAP0024340. The Caltech EPR Facility acknowledges the Beckman Institute and Dow Next Generation Educator Fund. We thank, in particular, the experimental support from beamline scientist Dr. Luke Daemen at ORNL and from Caltech's EPR facility manager Dr. Paul Oyala. Financial support from the U.S. Department of Energy (DOE), Office of Basic Energy Sciences, Quantum Information Science program (DE-SC0022920) is gratefully acknowledged.